\documentclass[aps,pra]{revtex4}

\usepackage{epsfig}
\usepackage{amssymb}
\usepackage{amsmath}
\usepackage{amsfonts}
\usepackage{graphicx}

\begin{document}
\title{Influence of slab thickness on the Casimir force}
\author{I. Pirozhenko\dag\ddag}
\author{A. Lambrecht\dag}
\affiliation{\dag Laboratoire Kastler Brossel,
CNRS, ENS, UPMC - Campus Jussieu case 74, 75252 Paris, France \\
\ddag Bogoliubov Laboratory of Theoretical Physics,
JINR, 141980 Dubna, Russia}
\pacs{42.50.Ct,12.20.Ds,12.20.-m}

\begin{abstract}
We calculate the Casimir force between slabs of finite thickness made of intrinsic
and doped silicon with different concentration of carriers and compare the results
to those obtained for gold slabs. We use the Drude and the plasma models to describe
the dielectric function for the carriers in doped Si. We discuss the possibility of
experimentally testing the appropriateness of these models. We also investigate the
influence of finite thickness on $VO_2$, which has recently been proposed for Casimir effect
measurements testing the metal-insulator transition.
\end{abstract}
\maketitle

\section{Introduction}
The availability of experimental set-ups that allow accurate
measurements of surface forces between macroscopic objects at
submicron separations has recently stimulated a renewed interest in
the Casimir effect \cite{Casimir} and  its possible applications
to micro- and nanotechnology.

In 1948  H. Casimir calculated the force between two plane-parallel mirrors
placed in vacuum at a distance $L$ apart from each other  and with the
area $A$ of the mirrors being much larger than the squared distance
$\left( A\gg L^{2}\right)$. In the  ideal case of perfectly
reflecting mirrors the force is given by  the following expression
\begin{equation}
F_{\rm Cas}= \frac{\hbar c\pi ^2 A}{240 L^4} \label{Casimir}
\end{equation}
with a positive value of $F_{\rm Cas}$
corresponding to  attraction, and a subsequent
negative pressure.

The Casimir force can be understood as the effect of radiation
pressure on the Fabry-Perot cavity formed by the two plane and
parallel mirrors. The intracavity vacuum energy being either
enhanced or suppressed, depending on whether the field frequency is
resonant or antiresonant, the net Casimir force results from the
balance between the repulsive and attractive contributions
associated respectively with these frequencies. The force is then
obtained as an integral over the axis of real frequencies, including
the contribution of evanescent waves besides that of ordinary waves,
and transformed into an integral over imaginary frequencies by using
physical properties fulfilled by real mirrors used in experiments
which do not require any further \textrm{adhoc} hypothesis.

Compared to the ideal situation considered by Casimir, a number of
corrections have to be taken into account in the calculation of the
force in real experiments. A large number of papers have been
devoted to the study of these effects and we refer the reader to
\cite{Bordag01} for an extensive bibliography. Here we will only be
concerned by the influence of the material properties and slab
thickness.

Considerable experimental progress has been achieved \cite{Bordag01}
in the control of the Casimir effect, opening the way to
applications in nano-science \cite{roukes,capasso}, particularly in
the development of nano- or microelectromechanical devices (NEMS or
MEMS). NEMS are movable nano mechanical structures inspired from
MEMS, with minimal critical dimensions of a few tens to a few
hundreds nanometers. NEMS approach constitutes a real technological
breakthrough to prepare the future generations of sensors and
actuators. At such small distances between the different elements,
the Casimir force in these systems may become quite important. It may,
on one hand, perturb the systems and produce stiction and adhesion
\cite{roukes}, but also be put to good use as an external force
allowing to change the systems' resonance frequency or introduce
bistable behavior \cite{capasso}.

In the past 10 years the Casimir effect has been studied extensively
for metals. Experiments have been performed for Au, Al, Cu in
different geometries and experimental set-ups. On the theoretical
side, calculations have taken into account the finite conductivity and frequency
dependent reflection coefficient of the different metals, modeling
them either by  plasma or Drude model or taking into account
tabulated optical data\cite{Lambrecht00,Genet00,Bordag01,Genet02,Jaekel91,Pirozhenko06}.
Only recently, a Casimir force measurement for Silicon bulk mirrors
has been reported on \cite{Mohideen06,Mohideen07} and calculations of the
Casimir force for Si bulks and slabs \cite{Esquivel03,AndreucciPLA}
have been performed however without investigating further the
observed difference in the bulk and slab behavior. The temperature
dependence of the force between Si plates has been studied in
\cite{Inui03}, while the influence of skin depth on the Casimir
force between metallic surfaces has been observed recently by
Capasso and collaborators~\cite{PNAS}.

The reference material in nano- or micro-electromechanical devices
is of course Silicon. In this paper, we present calculations of the
Casimir force between slabs of intrinsic or doped Silicon, Gold and
VO$_2$. In particular, we concentrate on the influence of the slab
thickness on the value of the force. We use the Drude and the plasma
model for the dielectric function of conductors and a Drude-Lorentz
model for Si and the insulating VO$_2$. We have previously shown
that for intrinsic Silicon the Casimir force depends strongly on the
slab thickness and  reduces considerably if the slab separation
exceeds the slab thickness, while for Gold force reduction is negligible,
except for very thin slabs~\cite{LPDA}. If the conductor is modeled by a plasma model
($\gamma=0$) the Casimir force becomes independent of the slab
thickness. Our present calculations show that for doped Silicon, the
Casimir force is diminished only in some distance range and for low
carrier levels and re-increases at long distances. We explain these
effects by analyzing the phase factor acquired by the vacuum field
due to finite slab thickness and the different behavior of the
dielectric functions. We also discuss the effect of a Gold
coating on the Si slab. Even though, a description of a thin Gold
film by a local dielectric function is not completely reliable
anymore, it gives some hint about the expected behavior. The slightest
Gold coating on intrinsic Silicon, even of only 1nm thickness,
re-increases the Casimir force between the two Si slabs. The above
described decrease in the force due to finite thickness of the Si
slab vanished because of the Gold coating.

We finally analyze the influence of slab thickness for VO$_2$ which has the particularity to
undergo a metal-insulator transition at 340 K~\cite{Verleur1968,Dachuan}. An experiment was proposed
recently~\cite{Mohideen07a} to investigate in detail the temperature dependence of the
Casimir force between gold and VO$_2$ film. The temperature dependence of the Casimir effect is
still an issue of controversial discussions \cite{Bostrom00,Svetovoy00,Bordag00, Lamoreaux01c,Sernelius01r,Sernelius01c,%
Bordag01r,Klimchitskaya01,Hoye03,Brevik05, Bezerra05,Buenzli05,Jancovici05,Brevik06,Brevik07}.
Here we make account only of the crucial change in the conductivity due to the phase transition,
but do not consider the temperature corrections to the Casimir effect itself.

\section{Formulation of the Casimir force between slabs of finite thickness}
The Casimir force is usually written as an integral over imaginary
frequencies and wavevectors \cite{Lifshitz1}. In order to visualize
easily the variation of the real Casimir force with respect to the
ideal formula (\ref{Casimir}) it is convenient to introduce  a
reduction factor $\eta_F=F/F_{Cas}$  \cite{Lambrecht00}
\begin{eqnarray}
&&\eta _{F}=\frac{120 L^4}{\pi ^{4}c}\sum_{p=\bot,||}
\int_{0}^{\infty }{\rm d}k k\int_{0}^{\infty}{\rm d}\omega \kappa \,
\emph{f}(\omega,k), \label{etaF}\\
&&\emph{f}(\omega,k)=\frac{r_{p}^{2}}{e^{2 \kappa L}-r_{p}^{2}},\quad
\kappa=\sqrt{k^2+\frac{\omega^2}{c^2}},\nonumber
\end{eqnarray}
$r_{p}$ denotes the reflection amplitudes  of the mirrors
at a given polarization $p$. This notation is a shorthand for
$r_{p}\left(i\omega ,i\kappa \right) $ where $i\omega $ is the
imaginary frequency and $i\kappa $ the imaginary wavevector along
the longitudinal direction of the cavity while $k$ is the modulus of
the transverse wavevector. The second integral is written over
imaginary frequencies as explained in detail in \cite{Lambrecht00}.

Assuming the plates to have a large optical thickness, the
reflection coefficients $r_{p}$ correspond to the ones of a mere
vacuum-metal interface \cite{LandauECM9}
\begin{eqnarray}
\rho_{\bot } &=&-\frac{\sqrt{\omega ^{2}\left( \varepsilon \left(
i\omega \right) -1\right) +c^{2}\kappa ^{2}}-c\kappa }{\sqrt{\omega
^{2}\left( \varepsilon \left( i\omega \right) -1\right) +c^{2}\kappa
^{2}}+c\kappa },
\nonumber \\
\rho_{||} &=&\frac{\sqrt{\omega ^{2}\left( \varepsilon \left(
i\omega \right)
-1\right) +c^{2}\kappa ^{2}}-c\kappa \varepsilon \left( i\omega \right) }{%
\sqrt{\omega ^{2}\left( \varepsilon \left( i\omega \right) -1\right)
+c^{2}\kappa ^{2}}+c\kappa \varepsilon \left( i\omega \right) },
\label{rThick}
\end{eqnarray}
$\rho_{p}$  stands for $\rho_{p}\left( i\omega ,i\kappa \right) $ and $%
\varepsilon \left( i\omega \right) $ is the dielectric constant of
the metal evaluated for imaginary frequencies.

However, for thin mirrors or slabs the
reflection coefficients depend on the  physical thickness $D$ and evaluate to
\cite{LambrechtPLA97,Genet03,PNAS,Lambrecht06}
\begin{eqnarray}
r_p &=&\rho_p \frac{1-e^{-2\delta }}{1-\rho_p ^{2}e^{-2\delta }},  \nonumber \\
\delta &=&\frac{D}{c}\sqrt{\omega ^{2}\left( \varepsilon \left(
i\omega \right) -1\right) +c^{2}\kappa ^{2}},  \label{rSlab}
\end{eqnarray}
$\varepsilon \left( i\omega \right) $ is the dielectric function of
the material evaluated for imaginary frequencies. $\delta $ is the
optical length of the slab. The single interface expression is
recovered in the limit of a large optical thickness $\delta \gg 1$.

\begin{figure}[t]
\epsfig{file=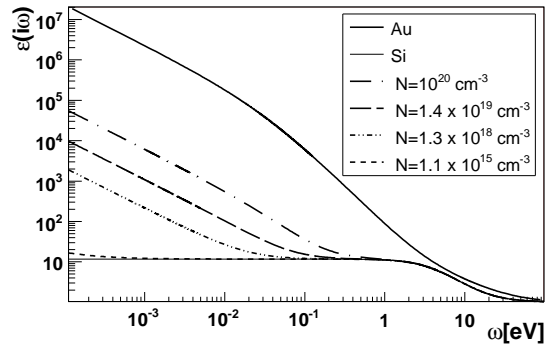,width=8cm} \caption{The different
dielectric functions for intrinsic and doped Silicon for varying
carrier densities in comparison with Gold.} \label{epsilonSi}
\end{figure}

The dielectric response function for real frequencies may be written
in terms of real and imaginary parts $\varepsilon ^{\prime }$ and
$\varepsilon ^{\prime \prime }$ obeying usual causality relations
which allow one to obtain the dielectric constant at imaginary
frequencies $\varepsilon \left( i\omega \right) $ from the function
$\varepsilon ^{\prime \prime }\left( x\right) $ evaluated at real
frequencies $x$ \cite{LandauECM9}
\begin{equation}
\varepsilon \left( i\omega \right) -1=\frac{2}{\pi }\int_{0}^{\infty }\frac{%
x\varepsilon ^{\prime \prime }\left( x\right) }{x^{2}+\omega
^{2}}{\rm d}x. \label{epsIm}
\end{equation}
The tabulated optical data for the complex index of refraction for
Silicon and Gold can be found in \cite{Palik}, while the data for
VO$_2$ has been given in a dedicated paper \cite{Verleur1968}. For
Silicon and VO$_2$ the data covers the whole relevant frequency
range and, in contrast to most metals, no extrapolation procedure at
low frequencies is necessary. We will give frequencies either in
$e$V or in rad/s, using the equivalence 1~$e$V $=1.519\times
10^{15}$~rad/s.  Figure \ref{epsilonSi} shows the
dielectric permeability of doped Silicon with different carrier
concentrations as a function of imaginary frequencies, obtained by
using (\ref{epsIm}), which we will need for the calculation of the
reduction factor of the Casimir force (\ref{etaF}). While at
low-frequencies  the dielectric function of intrinsic Silicon
approaches a constant value $\varepsilon_0=11.87$, $\varepsilon(\omega)$ of doped
Si increases with increasing carrier concentration and
 behaves like that of diluted metal. With increasing
frequency, the dielectric function of intrinsic Silicon is nearly
constant up to about $10^{15}$ rad/s and falls off only  for high
frequencies above a cut-off frequency, $\omega_0 \approx 6.6 \cdot
10^{15}$ rad/s, towards its asymptotic value
$\varepsilon_{\infty}=1.035$. $\varepsilon(\omega)$ of doped Silicon
decreases rapidly with increasing frequency and shows the same
cut-off frequency as intrinsic Si.

The dielectric function of intrinsic Silicon can well be
approximated by the following Drude-Lorentz function~\cite{Bergstrom97}
\begin{eqnarray}
\varepsilon_\mathrm{Si}(i\omega) = \varepsilon_{\infty}+
\frac{(\varepsilon_{0}-\varepsilon_{\infty}) \omega_0^2}{\omega^2 +
\omega_0^2}\label{epsilonapprox}
\end{eqnarray}
with the numerical values as given above. The dielectric function of
doped Si contains an additional part, which is modeled with a
dielectric function given by a Drude model
\begin{equation}
\varepsilon_\mathrm{dop}(i\omega) =\varepsilon_\mathrm{Si}(i\omega)
+ \frac{\omega_{\mathrm{p}}^2}{\omega(\omega +\gamma)}.
\label{epsilonSiDop}
\end{equation}

The different values of the plasma frequency $\omega_{\mathrm{p}}$
and the relaxation rate $\gamma$ for various carrier densities are
given in Table~\ref{tab1}.

\begin{table}
\centering
\begin{tabular}{l||c|c|c}
N ($\mathrm{cm}^{-3}$)& $\omega _{\mathrm{p}}$ (eV) & $\gamma$(eV) &
$\rho$ ($\Omega$ cm)
\\ \hline\hline $1.1\times10^{15}$ & $0.0021$ &
$0.0078$ & 13 \\
$1.3\times 10^{18}$ & $0.0725$ & $0.0247$  & $3.5\times 10^{-2}$\\
$1.4\times10^{19}$ & $0.238$ & $0.0518$ &  $6.8\times 10^{-3}$ \\
$10^{20}$ & $0.636$ & $0.06529$ &$1.2\times 10^{-3}$
\end{tabular}
\caption{The values of plasma frequency and relaxation rate for
various carrier densities, $\omega_\mathrm{p}=\sqrt{N
e^2/(\varepsilon_0 m^*)}$, $\gamma=N e^2 \rho/m^*$, where $m^*=0.34 m_e$
is the effective mass of the holes, and $\rho$ is the resistivity~\cite{AndreucciPLA}.}\label{tab1}
\end{table}

% the same in rad/c
%\begin{table}
%\centering
%\begin{tabular}{l||c|c}
%N ($\mathrm{cm}^{-3}$)& $\omega _{p}$
%($\times10^{12}\mathrm{rad/s}$) & $\omega _{\tau }$
%($\times10^{13}\mathrm{rad/s}$)
%\\ \hline\hline $1.1\times10^{15}$ & $3.20$ &
%$1.182$  \\
%$1.3\times 10^{18}$ & $110.18$ & $3.760$  \\
%$1.4\times10^{19}$
%& $361.59$ & $7.688$  \\
%$10^{20}$ & $966.38$ & $9.918$
%\end{tabular}
%\caption{}\label{tab1}
%\end{table}
The description of the dielectric function of the p-doped Silicon by
the model~(\ref{epsilonSiDop}) is valid for the doping levels up to
$10^{20}$ cm$^{-3}$. For higher doping levels Si becomes
degenerated.

\begin{figure}[ptb]
\centerline{\psfig{figure=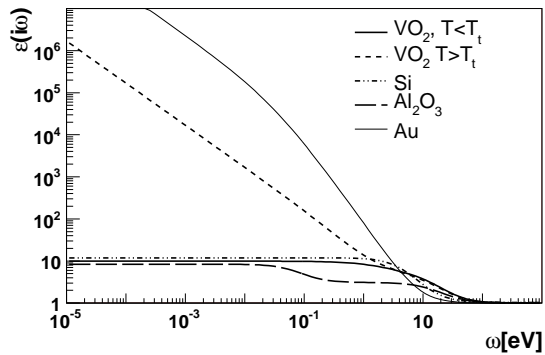,width=8cm}}
\caption{Dielectric function for VO$_2$ above and below the critical
temperature in comparison with the one of intrinsic Si, Gold and
Al$_2$O$_3$.} \label{epsilonVO2}
\end{figure}

The dielectric function of VO$_2$ is shown in Fig.~\ref{epsilonVO2} above and below the critical temperature $T_t$.
Below the critical temperature we used the model already applied in
\cite{Mohideen07a} which had been proposed first in
\cite{Verleur1968}
\begin{equation}
\varepsilon_n(i\omega)=1+\frac{\varepsilon_n(i\infty)-1}{1+\frac{\omega^2}{\omega_{\infty}^2}}+
\sum_{i=1}^{7}\frac{s_{n,i}}{1+\frac{\omega^2}{\omega_{n,i}^2}+
\Gamma_{n,i}\frac{\omega}{\omega_{n,i}}}
\label{VO2_eps}
\end{equation}
with $\varepsilon_{n}(i\infty)=4.26$, $\omega_{\infty}=15 eV$. For
the rest of the parameters see Table~\ref{tab2}. Above the critical
temperature the dielectric permeability is given by
\begin{eqnarray}
\varepsilon_{\tilde{n}}(i\omega)&=&1+\frac{\omega^2_{p,\tilde{n}}}{\omega(\omega+\gamma_{\tilde{n}})}+
\frac{\varepsilon_{\tilde{n}}(i\infty)-1}{1+\frac{\omega^2}{\omega_{\infty}^2}}\nonumber\\
&&+\sum_{i=1}^{4}\frac{s_{\tilde{n},i}}{1+\frac{\omega^2}{\omega_{\tilde{n},i}^2}+
\Gamma_{\tilde{n},i}\frac{\omega}{\omega_{\tilde{n},i}}}
\label{VO2_eps2}
\end{eqnarray}
where $\varepsilon_{\tilde{n}}(i\infty)=3.95$, $\omega_{\mathrm{p},\tilde{n}}=3.33$ eV, $\gamma_{\tilde{n}}=0.66$ eV.
\begin{table}
%\centering
\begin{tabular}{ccc}
$s_n$&$ \omega_n$(eV) & $\Gamma_n$  \\ \hline\hline
0.79& 1.02  & 0.55  \\
0.474 & 1.30 & 0.55  \\
0.483& 1.50 & 0.50  \\
0.536& 2.75 & 0.22  \\
1.316& 3.49 & 0.47  \\
1.060& 3.76 & 0.38  \\
0.99 & 5.1 &  0.385  \\
\hline
\end{tabular}\quad \qquad
\begin{tabular}{ccc}
$s_{\tilde{n}}$& $ \omega_{\tilde{n}}$(eV) & $\Gamma_{\tilde{n}}$ \\ \hline\hline
1.816& 0.86 &0.95   \\
0.972& 2.8 & 0.23  \\
1.04& 3.48 & 0.28 \\
1.05& 4.6 & 0.34  \\
\hline
\\
\\
\\
\end{tabular}
\caption{The values of the parameters in the models~(\ref{VO2_eps}) and
(\ref{VO2_eps2})~\cite{Verleur1968}}\label{tab2}
\end{table}

Clearly, above the critical temperature the material behaves more
like a metal with a strongly increasing dielectric function at low
frequencies. For low temperatures its dielectric function nearly
matches the one of intrinsic Si, as VO$_2$ becomes an insulator and
the dielectric function at zero frequency has a constant value of
about $\epsilon_n(i0) \sim 10$.

\section{Numerical results for doped Silicon}
We now evaluate the Casimir force (\ref{etaF}) using the tabulated
optical data for different slab thickness $D$. The numerical
procedures follow the principles described in \cite{Lambrecht00}.

\begin{figure}[ptb]
\centerline{\psfig{figure=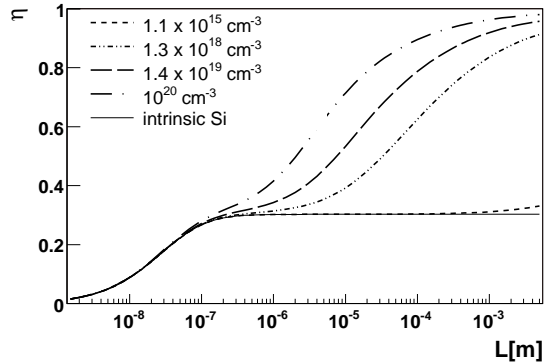,width=8cm}}\caption{Reduction
factor of the Casimir force between two bulk mirrors of intrinsic or
p-doped Silicon. } \label{Sibulk}
\end{figure}

\begin{figure}[ptb]
\centerline{\psfig{figure=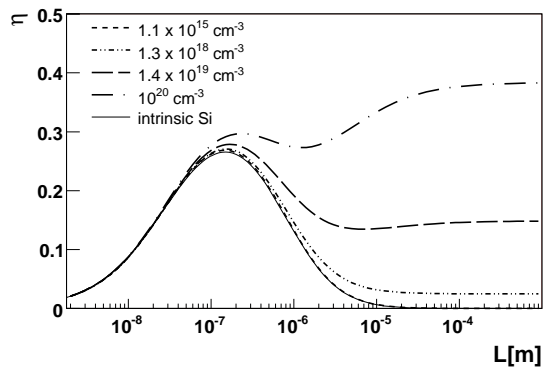,width=8cm}}\caption{Reduction
factor of the Casimir force between two 100nm thin slabs of
intrinsic or p-doped Silicon. } \label{Sislab100}
\end{figure}

Figures \ref{Sibulk} and  \ref{Sislab100} show $\eta _{F}$ as a function
of plate separation for intrinsic and p-doped Silicon for bulk mirrors
and for Si slabs of 100nm thickness. Between two bulk
mirrors the force reduction factor is a continuously growing
function of the mirror separation and reaches a constant value
$\eta_F(\infty) \approx 0.303$ in the long distance limit, which
means that $F \approx F_\mathrm{Cas}/3$. In contrast, between two
Silicon slabs the Casimir force reduction factor grows continuously
only for separations $L \simeq D$, while it starts to diminish
considerably when the slab separation becomes of the order of or
exceeds the physical slab thickness~\cite{LPDA}. This effect
disappears for doped Silicon. Only a small decrease in the force
reduction factor remains at intermediate distances (about 1$\mu$m).
For larger distances the conduction due to the carriers introduced by doping
re-increases the value of the reduction factor.

%\begin{figure}[ptb]
%\centerline{\psfig{figure=500nm_b.eps,width=8cm}}\caption{Reduction
%factor of the Casimir force between two 500nm thin slabs of
%intrinsic or doped Silicon. } \label{Sislab500}
%\end{figure}

This result could have very interesting consequences in
nanotechnology. The observed decrease in the Casimir force  for
intrinsic Si in the distance range $L>D$ can be canceled by
injecting carriers. This allows to tune  the Casimir force
within certain restrictions using at the same time the slab
thickness and the carrier densities as variable parameters.

In \cite{Mohideen07} the first experiment of optical modulation of
dispersion forces through the change of carrier density by laser
pulses was reported. The irradiation of a silicon slab by laser
pulses allows to achieve charge carrier concentrations of
$n=(2.0\pm0.4)\times10^{19}$ cm$^{-3}$. Here we present numerical
results for the Casimir force obtained by assuming that the
contribution  of the induced charge carriers is given by Drude  or
plasma terms,
$\sum_{i}\omega_{\mathrm{p},i}^2/(\omega(\omega+\gamma_{i}))$ or
 $\sum_{i}\omega_{\mathrm{p},i}^2/\omega^2$, $i=e,p$, with
  $\omega_{\mathrm{p},p}=0.368$  eV, $\gamma_{p}=0.00329$  eV  for the
  positive induced carriers and $\omega_{\mathrm{p},e}=0.329$ eV,
    $\gamma_{e}=0.01185$  eV for the negative ones.

\begin{figure}[pbb]
\psfig{figure=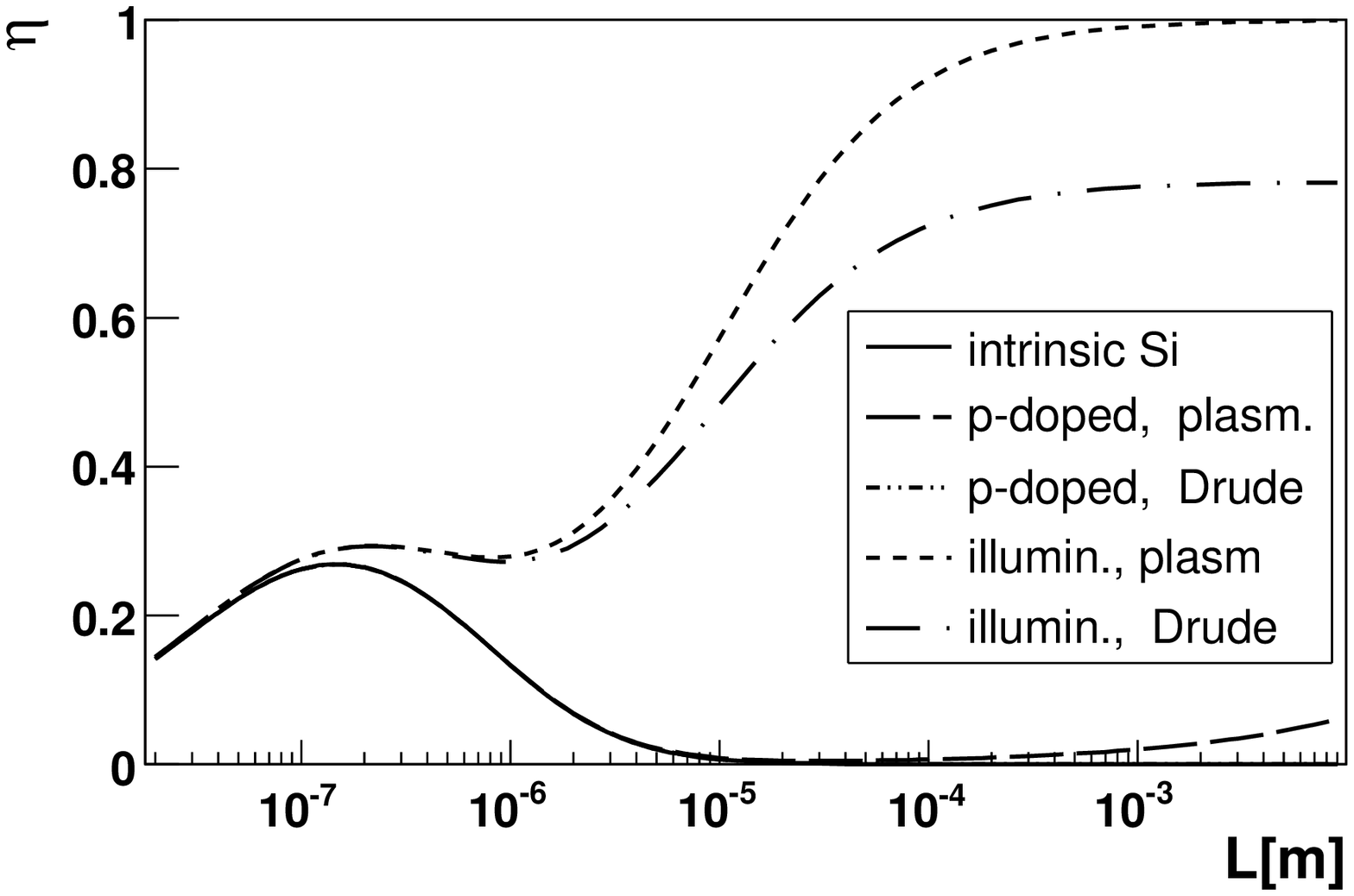,width=8cm}
\psfig{figure=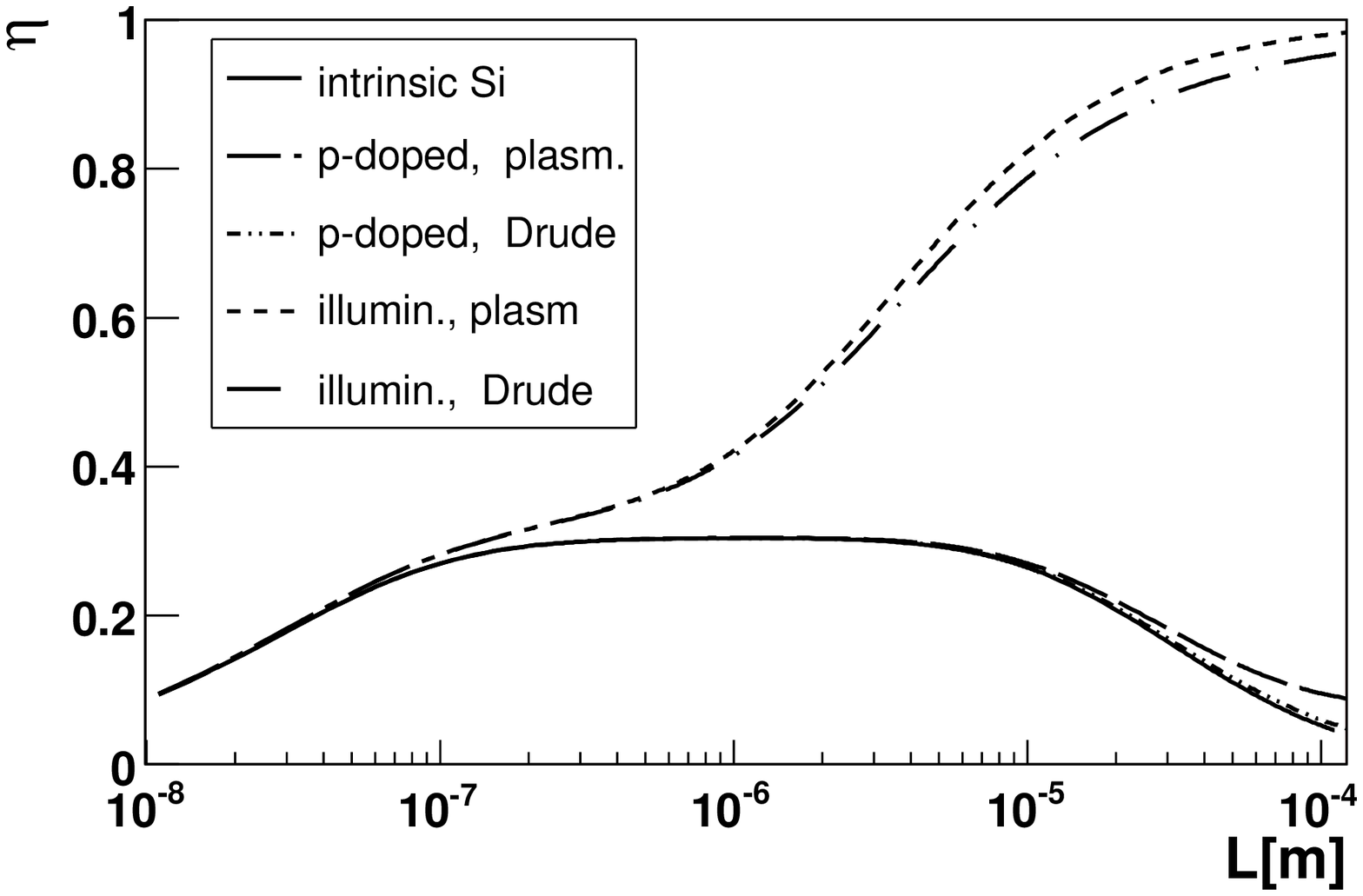,width=8cm}
\caption{Reduction
factor of the Casimir force between two 100nm (upper graph) or 4000nm slabs of
Silicon with the carrier density modulated by laser irradiation in
comparison with p-doped Silicon, $n=5\times10^{14}$ cm$^{-3}$,
$\omega_{\mathrm{p}}$=0.00184 eV,
    $\gamma$=0.00329 eV.} \label{Si_laser}
\end{figure}

Fig.~\ref{Si_laser} gives the numerical results for the reduction
factor of Casimir force between two silicon slabs irradiated by
laser pulses in comparison with p-doped silicon. Here we took the
parameters from the paper~\cite{Mohideen07}. The doping is described
by the Drude or plasma model. The difference between the models
could manifest itself at distances of several micrometers for highly
doped silicon provided the slab is thin (upper graph).
However the experiment requires a mirror thick enough for irradiation not to penetrate
into the cavity. The thickness should be greater than the optical absorption depth
of Si at the wavelength of the laser beam used for the doping.
If the thickness of the slab is 4000 nm as used in~\cite{Mohideen07} the experiment can
hardly  distinguish between  Drude and plasma models.

\section{Numerical results for VO$_2$}

\begin{figure}[ptb]
\centerline{\psfig{figure=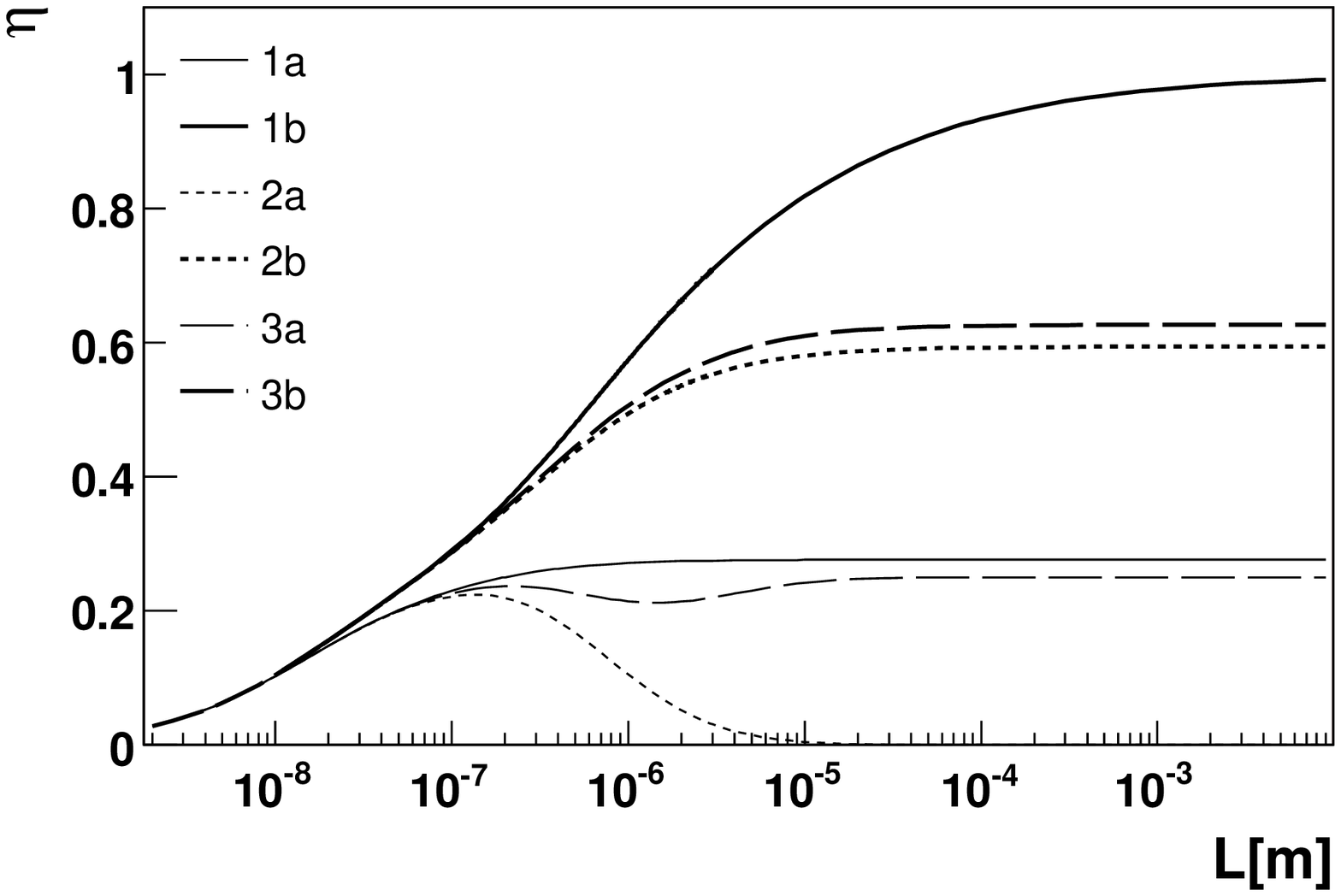,width=8cm}}\caption{Reduction
factor of the Casimir force between two VO$_2$ bulks (1), two 100~nm slabs of VO$_2$ (2), and
two 100~nm VO$_2$ layers on  Al$_2$O$_3$ bulk substrate (3)
below (a) and above (b)
the critical temperature. } \label{VO2}
\end{figure}

In~\cite{Mohideen07a} a compound mirror, having a 100nm layer of
VO$_2$ on an Al$_2$O$_3$ bulk  substrate, is treated as an effective
medium with a dielectric permittivity~(\ref{VO2_eps},\ref{VO2_eps2}).
The authors rely on the fact that the parameters in  Table~\ref{tab2}
were retrieved from  reflectivity and transmissivity spectra exactly for the same
two-layered system as in Ref.~\cite{Verleur1968}.

However,~\cite{Verleur1968} contains also the description of the fitting procedure
that uses the formulas corresponding to two layered system. The
finite thickness of the substrate is neglected in the analysis,
but the account is made for the finite thickness of the VO$_2$ film.
The dielectric function of the substrate (sapphire) is considered as known.
It is substituted into the formulas, and the unknown parameters of the  model
describing the VO$_2$ film are found by the curve fitting programme.
Since it is not quite clear from~\cite{Verleur1968} if the
parameters in the Table~\ref{tab2} and the model  correspond to the effective media
or to the $VO_2$ itself,  we considered both possibilities.

First we calculated the reduction factor for the Casimir force between two "bulks"
of the effective medium. In other words we use the dielectric permittivities~(\ref{VO2_eps})
and~(\ref{VO2_eps2}) with the parameters from Table~\ref{tab2} and formulas for bulk reflection coefficients.
The corresponding curves in Fig. \ref{VO2} are plotted  by thin full line below
the phase transition, $T<T_t$, and by thick full line above it, $T>T_t$.

The reduction factor for the Casimir force between two 100 nm VO$_2$
slabs  below the phase transition  is given by thin dotted line.
We see that before  the phase transition the behavior of
VO$_2$ slabs resembles the one of the silicon slabs. Above the phase
transition ($T>T_t$, thick dotted line) the material behaves like a
dilute metal, and  we do not observe the decrease of the reduction factor
at $L>200$nm. However the reduction factor for a  slab never reaches
the unit value it does for pure metals at zero temperature in the
long distance limit.

We also calculate the reduction factor for  $100$nm VO$_2$ film on a
sapphire substrate. It means that we used the formulae for two-layered mirrors.
The layer facing the cavity has the dielectric permittivity
(\ref{VO2_eps}) at $T<T_t$ and (\ref{VO2_eps2}) at $T>T_t$. For the substrate
we used the dielectric permittivity of Al$_2$O$_3$
$$\varepsilon_{Al_2O_3}(i\omega)=1+\frac{A_1}{1+\omega^2/f_1^2}+\frac{A_2}{1+\omega^2/f_2^2}+\frac{A_3}{1+\omega^2/f_3^2}$$
with $A_1=1.023$, $A_2=1.058264$, $A_3=5.280792$, and
$f_1=20.19$ eV, $f_2=11.21$ eV, $f_3=0.07$ eV~\cite{mellesgriot}.

Below the transition temperature the dielectric
function of VO$_2$ is close to the one of Al$_2$O$_3$ (see Fig.~\ref{epsilonVO2}).
That is why the result for the Casimir force
between two effective media bulks almost coincides with the force
between two-layered mirrors, VO$_2$ on sapphire substrate. When the
temperature overpasses the critical value, the entire bulk of the effective media
becomes a metal, while in the case of two-layered mirror, the
substrate remains an insulator. Consequently, at large distances the
force between two effective media bulks  is larger than the force between
compound mirrors. The force obtained within the effective media calculation  differs  from the
result of the two-layer calculation starting from the  distance of several hundred
nanometers (Fig.~\ref{VO2}).

\begin{figure}[pbb]
\centerline{\psfig{figure=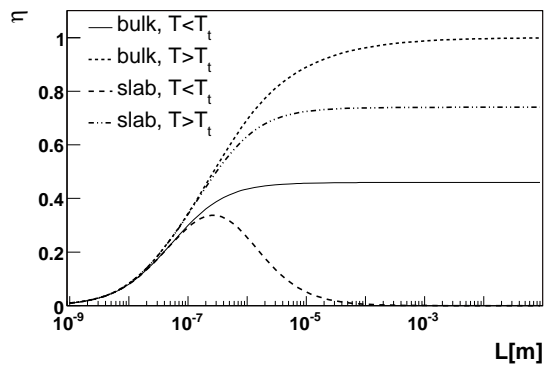,width=8cm}}\caption{Reduction
factor of the Casimir force between  VO$_2$ and Au mirrors above and
below the critical temperature. } \label{VO2vrsAu}
\end{figure}

Finally in Fig. \ref{VO2vrsAu} we present the results for the VO$_2$ mirror in front of the
Gold mirror. The curves obey the same
distance dependencies as in the case of equal mirrors. But the
values of the reduction factor are increased thanks to the high
reflectivity of gold. This setup is favorable from the experimental
point of view.

It is important to note that the dielectric function we use corresponds to thin film
measurements. If the force is measured between true bulk VO$_2$ mirrors, the results
should be compared with the calculation that uses the bulk dielectric function~\cite{Verleur1968} .

\section{Discussion and Conclusion}
The obtained results can be understood in terms of the optical
length of the layer material or the phase factor acquired by the
field while propagating through the finite layer.

\begin{figure}[ptb]
\centerline{\psfig{figure=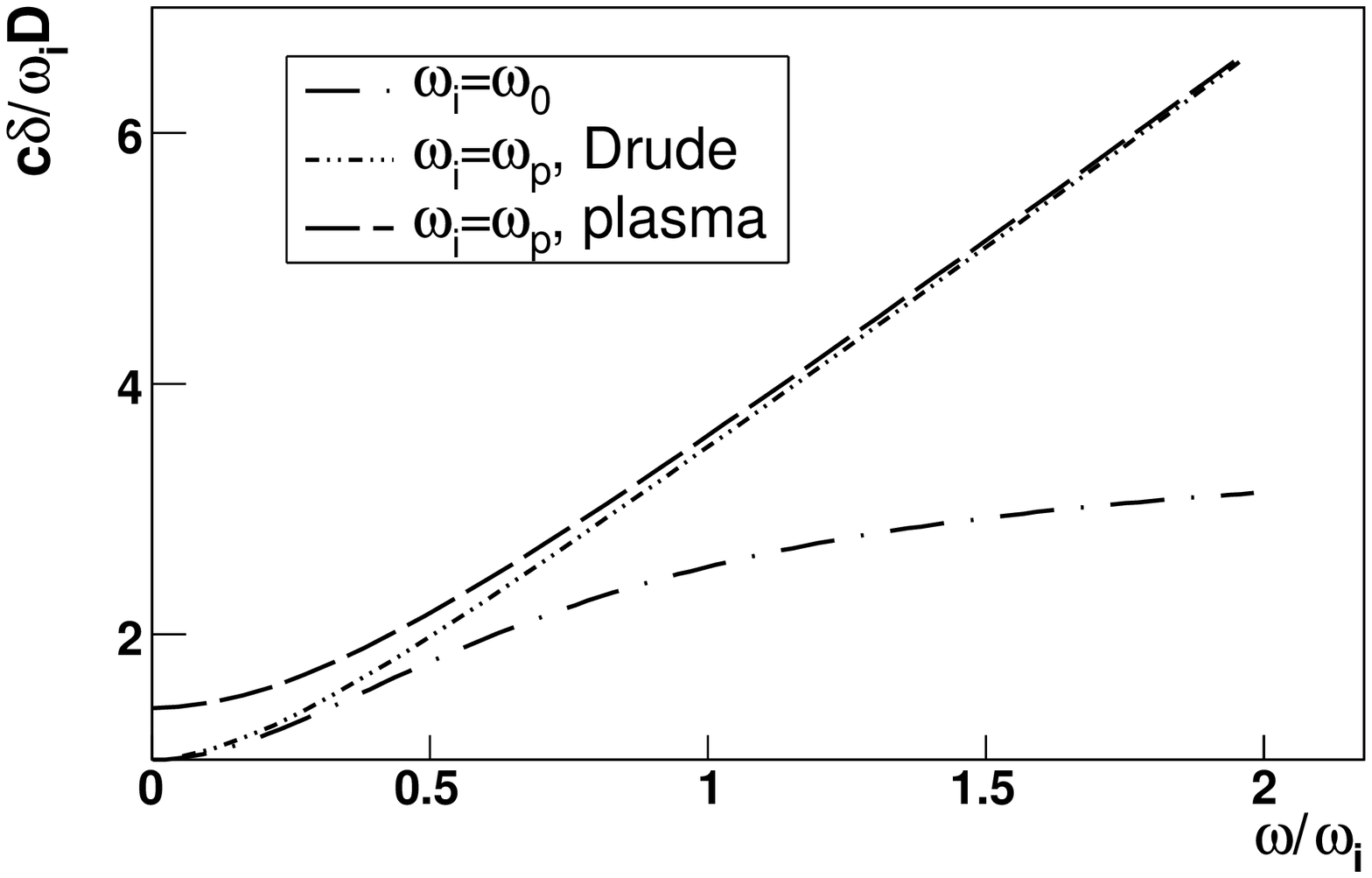,width=8cm}}\caption{Phase factors acquired by the
vacuum field when it propagates through the intrinsic and weakly p-doped silicon, $c\kappa/\omega_i=1$,
$n=5\cdot10^{14}$cm$^{-3}$, $\omega_p^{(p)}=1.84\cdot10^{-3}$eV,
$\gamma^{(p)}=3.29\cdot10^{-3}$eV,~\cite{Mohideen07a}.} \label{Mohideen_pf1}
\end{figure}

%\begin{figure}[ptb]
%\centerline{\psfig{figure=phfact_Si_doped_Mohideen2.eps,angle=90,width=8cm}}\caption{Phase factor, Mohideen2} \label{Mohideen_pf2}
%\end{figure}

\begin{figure}[ptb]
\centerline{\psfig{figure=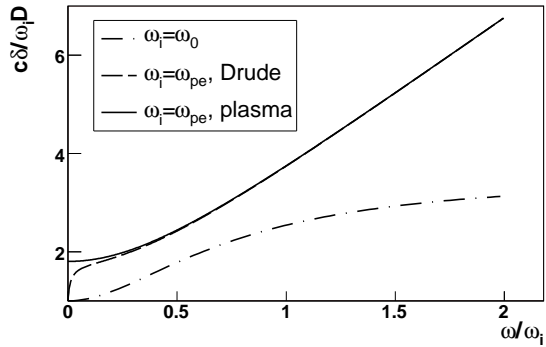,width=8cm}}\caption{Phase factors acquired by the
vacuum field when it propagates through the intrinsic and laser illuminated silicon,
$c\kappa/\omega_i=1$, $n=(2.0\pm0.4)\times10^{19}$ cm$^{-3}$}
\label{Mohideen_pf3}
\end{figure}

As indicated in~(\ref{rSlab}), the phase factor $\delta$ is
frequency-, wavevector- and thickness-dependent. Using the
dielectric function of Silicon, which is typical for insulators, in expression
(\ref{rSlab}) we find
\begin{equation}
\delta_\textrm{ins} =\frac{D}{c}\sqrt{\omega ^{2}\left(
\varepsilon_{\infty} -1 +
\frac{(\varepsilon_0-\varepsilon_{\infty})\,\omega_0^2}{\omega^2
+\omega_0^2}\right) +c^{2}\kappa ^{2}}. \label{deltaSi}
\end{equation}
The phase factor (\ref{rSlab}) for metals or doped Si slabs modeled
by a Drude or plasma model ($\gamma=0$) evaluates to
\begin{equation}
\delta_\mathrm{Drude}
=\frac{D}{c}\sqrt{\omega_{\mathrm{p}}^2\frac{\omega}{\omega+\gamma}
+c^{2}\kappa ^{2}}, \quad
\delta_{\textrm{plasma}}=\frac{D}{c}\sqrt{\omega_{\mathrm{p}}^2
+c^{2}\kappa ^{2}}. \label{deltaAu}
\end{equation}
Let us now compare these phase factors at some constant fixed value
for $\kappa$, and $D$. Fig.~\ref{Mohideen_pf1} shows the phase
acquired by the vacuum field while propagating through the intrinsic
and doped silicon. The doping is described using either a Drude
(dashed-double-dotted curve) or a plasma (dashed curve) model.

The phase factors of  intrinsic  and weakly doped silicon
are indistinguishable at low frequencies, provided that the doping is
described by Drude model, Fig.~\ref{Mohideen_pf1}.
The phase factors acquired by the field in the silicon slab previously illuminated by laser,
Fig.~\ref{Mohideen_pf3}, coincide in Drude and plasma model
descriptions down to the values of the ratio $\omega/\omega_{pe}\sim0.5$. For lower frequencies
the plasma model curve approaches the value $\sqrt{2}$, and the Drude curve approaches zero.
Though the difference  between Drude and plasma curves reduces as the carrier
density is increased, their behavior is  qualitatively different in $\omega,\kappa\to0$ limit.
It results in different predictions for the force at long distances.

The largest contribution to the Casimir force
comes from the frequencies around the characteristic
frequency $\omega_{ch}\sim c/L$. Large plate separation thus
corresponds to small frequencies and small wavevectors $\kappa$.
At small frequencies and small wave-vectors the
bulk reflection coefficients~(\ref{rThick}) tend to their static value
\begin{equation}
\lim_{\omega,\kappa\to0}\rho_{\bot,\|}=
\frac{1-\sqrt{\varepsilon(i0)}}{1+\sqrt{\varepsilon(i0)}}
\label{static_refl}
\end{equation}
which is $-0.55$ for intrisic silicon, $-0.52$ for VO$_2$ before the phase transition, and  $-1$ for gold or any metal
described by Drude or plasma model. Substituting these values into the integral
(\ref{etaF}), or in other words neglecting the  dispersion at large separations
of the mirrors,  we get rough estimation for $\eta^{Si}_{\infty}\simeq 0.28$
for two silicon bulks, $\eta^{V0_2}_{\infty}\simeq 0.25$ for two V0$_2$ bulks,
and $\eta^{met}_{\infty}\simeq 1$ for two metallic ones.

In~\cite{LPDA} we have shown that the effect of finite slab
thickness manifests itself for Silicon at plate separations of
the order $L>c/\omega_0$, with $\omega_0=6.6 \cdot10^{15}$ rad/s.

The optical length of a silicon slab~(\ref{deltaSi}) at large plate separations tends
to zero as $\delta_{ins}= D/c\sqrt{\varepsilon(i0)-1}\;\omega+O(\omega^3)$, while
the bulk reflection coefficient goes to its  static value~(\ref{static_refl}).
The numerator of~(\ref{rSlab}) vanish. Consequently the reflection coefficients of thin slabs
at large plate separations vanish too. The force is considerably reduced.

The optical length of a metallic slab~(\ref{deltaSi}) at small frequencies and small transversal
wave-vectors  tends to a constant for the plasma model, $\delta^{pl}=D\omega_p/c$, while the bulk
reflection coefficients tend to $-1$ for both field polarizations. The reflection coefficient of a
slab described by plasma model approaches its bulk value. The long distance limit is then not
affected by the slab thickness except for very thin slabs.
For gold with $\omega_p=9 eV$, two 10nm slabs separated by $L=10^{-4}$m
yield up to the third
decimal sign $\eta=F/F_C=0.997$. The reduction factor corresponding
to 50 or 100nm slabs is $\eta=0.999$ which coincides with the one for two bulks
at the same separation.

For Drude model the optical  length tends to zero as
$ \delta^{Dr}= D\omega_p/(c \sqrt{\gamma})\sqrt{\omega}+O(\omega^{3/2})$,
while the bulk reflection coefficient for  metals tends to
$\rho=-1+2\sqrt{\gamma}\sqrt{\omega}/\omega_p+O(\omega)$.
The reflection coefficient is then  $r\sim 2\rho\,\delta/(1-\rho^2(1-2\delta))$.
Substituting here the low frequency expansions for $\rho$ and $\delta$
and confining ourselves to the lowest power of $\omega$ we get
\begin{equation}
r\sim-\frac{1}{1+\frac{\Lambda}{D}}, \quad \Lambda=2\gamma c/\omega_p^2.
\label{refl_slab}
\end{equation}
The absolute value of the slab reflection coefficient~(\ref{refl_slab}) is essentially
smaller than one if $D\leq\Lambda$. For gold with $\omega_p=9 eV$,
$\gamma=0.035$eV, the effective thickness is $\Lambda=1.7\times10^{-10}$m.
Therefore in the long separation limit for slabs as thin as $20$nm  the reflection coefficients
mount to $-0.991$! For two bulk mirrors separated by the distance $L=10^{-4}$m the reduction
factor is $0.993$. The slab reduction factor is lower than the bulk one at the same separation,
but the effect is weak in comparison to dielectric mirrors. With the separation between
plates $L=10^{-4}$m, the reduction factors are $0.922$,  $0.955$, and $0.978$ respectively for
10, 20 and 50 nm slabs.

We conclude that it is enough to coat a silicon slab with a few nanometer
layer of metal to suppress the effect of finite slab thickness.

Though at low frequencies the phase factor acquired by the field in doped Si behaves similarly
to the one in gold~ (Fig.~\ref{Mohideen_pf1},~\ref{Mohideen_pf3}),
the effective thickness $\Lambda$
is much lower for gold  than for  doped Silicon where it varies from
$7\times10^{-4}$m  for the doping level of $ 10^{15}$
cm$^{-3}$ to $64$nm  for $\sim 10^{20}$ cm$^{-3}$. That is why we observe considerable reduction of the
force already for 500 nm slabs. For 100 nm slab of doped silicon, $n=10^{20}$ cm$^{-3}$,
we get $r\sim -0.61$, and $\eta\sim0.35$. This rough estimation is in good agreement with our
numerical result, $\eta\sim0.38$, see Fig.~\ref{Sislab100}.

In section IV we presented the numerical results for VO$_2$ which behaves as a semiconductor
at temperatures below $T_t=340$K and as a metal at higher temperatures.
In the semiconductor state of this material the force between finite slabs reproduces the behavior
characteristic for silicon. After the phase transition the slabs of VO$_2$
are attracted as Drude metals.

Of course all preceding results are based on a number
of models, in particular the plasma and Drude models. As the issue of
the precise description of finite conductivity for metals and
semiconductors is still not satisfactorily settled, it would be very
interesting to test the previously presented results experimentally.
This might help to clarify the open problem of the temperature
corrections to the Casimir force \cite{Brevik,Onofrio,Mostepanenko,Sernelius,Esquivel}.

\section{Acknowledgements} We acknowledge financial support
from the European contract STRP 12142 NANOCASE.

\end{document}